\documentclass[aps,prl,twocolumn,showpacs,amsmath,amssymb,superscriptaddress]{revtex4}

\usepackage{graphicx}
\usepackage{amsmath}
\usepackage{amssymb}
\usepackage{color}

\begin{document}

\title{Understanding water's anomalies with locally favored structures}
\author{John Russo}
\affiliation{Institute of Industrial Science, University of Tokyo, 4-6-1 Komaba, Meguro-ku, Tokyo 153-8505, Japan}
\author{Hajime Tanaka}
\thanks{tanaka@iis.u-tokyo.ac.jp}
\affiliation{Institute of Industrial Science, University of Tokyo, 4-6-1 Komaba, Meguro-ku, Tokyo 153-8505, Japan}

\date{\today}

\begin{abstract}
Water is a complex structured liquid of hydrogen-bonded molecules that displays a surprising array of unusual properties,
also known as water anomalies, the most famous being the density maximum at about $4^\circ$C. 
The origin of these anomalies is still a matter of debate, and so far
a quantitative description of water's phase behavior starting from the molecular arrangements 
is still missing. Here we provide its simple physical description from microscopic data obtained through computer simulations.
We introduce a novel structural order parameter, which quantifies the degree of translational order of the second shell, and show that 
this parameter alone, which measures the amount of locally favored structures, accurately characterizes the state of water.
A two-state modeling of these microscopic structures is used to describe the behavior of liquid water
over a wide region of the phase diagram, correctly identifying the density and compressibility anomalies,
and being compatible with the existence of a second critical point in the deeply supercooled region.
Furthermore, we reveal that locally favored structures in water not only have translational order in the second shell, but also contain five-membered rings of hydrogen-bonded molecules.
This suggests their mixed character: the former helps crystallization, whereas the latter causes frustration against crystallization.
\end{abstract}

\maketitle

\section*{Introduction}

The anomalous thermodynamic and kinetic behavior of water is known to play a fundamental role not only in many physical
and chemical processes in materials science, but also in
biological, geological and terrestrial processes in nature~\cite{Eisenberg,AngellR,mishima_stanley,debenedetti_review}.
For this reason, a lot of effort has been
devoted to rationalizing water's anomalous behavior in a coherent and simple physical picture,
but no consensus has yet emerged. One of the breakthroughs in this endeavor was the discovery of
water's polyamorphism, i.e. the existence of amorphous coexisting phases in the supercooled region
of the phase diagram. Distinct states
have indeed been found in glassy water, called low-density (LDA), high-density (HDA) and very high-density (VHDA)~\cite{loerting2011many}
amorphous ices, which can interconvert with each other by the application of pressure.
It is believed that the transition between the amorphous ices connects to a liquid-liquid
first-order phase transition line above the glass transition temperature ($T_g$), and terminates
at a critical point~\cite{poole_nature}, but the fundamental nature of this transition is still
being debated~\cite{limmer_chandler}.
The verification of the liquid-liquid critical point scenario (LLCP) is hindered by water's crystallization
at large supercooling~\cite{stokely_cooperativity}, so that much of the evidence comes from computer simulation studies~\cite{poole_nature,liu2009low,gallo_ising,poole_st2},
and only indirectly from experiments where crystallization is suppressed either by strong spatial confinement~\cite{Mallamace}
or by mixing an anti-freezing component~\cite{murata}.

One way to understand water's polyamorphic behavior is to introduce the concept of locally
favored structures \cite{tanaka2000thermodynamic,tanaka2000simple,tanaka_review}, which are defined as particular long-lived molecular arrangements which correspond
to some local minima of the free energy. In this view, water's polyamorphism comes from the competition
between two different types of molecular arrangements~\cite{mishima_stanley}: one in which the different tetrahedral units form open
structures, and the other with a smaller specific volume due to a high degree of interpenetration.
The presence of two amorphous fluid phases has indeed been observed in
computer simulations of some models of water, where the freezing transition can be avoided~\cite{poole_st2,kesselring2012nanoscale,liu2012liquid}.
But whether locally favored states exist above the critical region is still a matter of debate.
Many physical quantities exhibit behaviors suggestive of two states in liquid water, 
such as infrared and Raman spectra \cite{Eisenberg}, and
the presence of an isosbectic point in Raman spectra has been regarded as a clear indication supporting a mixture model 
since its finding by Walrafen \cite{Walrafen_T}.  
Some evidence for the inhomogeneous structure of water was also reported in experimental studies of X-ray absorption
spectroscopy, X-ray emission spectroscopy and X-ray small angle scattering~\cite{huang2009inhomogeneous,nilsson2011perspective},
but these results are highly debated~\cite{soper_gordon,saykally_gordon}, 
especially since the majority component at room temperature was proposed to be associated with the break-up of the tetrahedral structure. 
Despite these pieces of evidence supporting a two-state picture, the lack of a clear connection between these experimental observables and the amount of locally favored structures
has made it difficult to estimate the fractions of the two states in a convincing manner.
From a phenomenological standpoint, two-state models have been extremely successful in describing the anomalies of water
using a restricted number of fitting parameters~\cite{tanaka2000thermodynamic,tanaka2000simple,tanaka_review,holten_anisimov},
and recent simulations by Cuthbertson and Poole~\cite{poole_mixture} have opened the way for a quantitative assessment of these models
from miscroscopic informations, but only for state points around the Widom line, i.e., at temperatures 
not accessible to experiments and far from the anomalies.
The essential difficulty in defining locally favored states is finding a structural order parameter
that directly correlates with water's anomalies. Several attempts have been made,
each differing in the microscopic
definition of the states involved. Examples include states based on ice polymorphs~\cite{robinson},
tetrahedral order~\cite{Errington,nilsson2011perspective,limmer_molinero}, or relative distance between neighbors~\cite{appignanesi,poole_mixture,wikfeldt_bimodal}.

To overcome this difficulty, we introduce a new structural order parameter that quantifies the degree of translational order in the second shell.
We show that, for one of the most reliable computer models, water anomalies are a consequence of
translational ordering of the second shell and can be described very well by a two-state model.
We further identify the structural characteristics of the locally favored structures of water
and their link to the unusually large degree of supercooling of the liquid phase.

\section*{Results and Discussion}
\subsection*{Relevant structural order parameter. }
First we explain a key idea to identify the relevant structural order parameter for characterizing the water structure.
Our locally favored structures correspond to local structures with low energies $E$, high specific volumes $v$ and low degeneracy $g$. In the following we denote 
these structures as the $S$ state. In contrast, thermally excited states are characterized by a high degree of disorder and degeneracy, low specific volumes
and high energies. We label these structures as the $\rho$ state. In formulas, $v_\rho<v_S$, $E_S<E_\rho$ and $g_S\ll g_\rho$.
To identify the $S$ state we introduce a new order parameter $\zeta$ which measures local
translational order in the second shell of neighbors.
The importance of the second shell structure was notably pointed out by Soper and Ricci \cite{soper2000}. 
Translational order is a measure of the relative spacing between neighboring particles,
and it is one of the fundamental symmetries broken at the liquid-to-solid transition~\cite{russo_hs}. Locally, a molecule is in a
state of high translational order if the radial distribution of its neighbors is ordered. A liquid, by definition,
cannot have full translational order, but it might display translational order on shorter scales.
Water molecules, even at high temperatures, displays a high level of tetrahedral symmetry, meaning that a high
degree of translational order is always present up to the first shell of nearest neighbors. 
In this sense, tetrahedral order itself is not enough to describe water's anomalies. 

To determine locally favored states, we focus instead on ``translational order of second nearest neighbors''. 
The operational definition goes as
follows (a schematic representation is given in Fig.~\ref{fig:1}A). For water molecule $i$ (labeled \textbf{0} in Fig.~\ref{fig:1}A) we order its neighbors
according to the radial distance $d_{ji}$ of the oxygen atoms; the order parameter $\zeta(i)$ is
then defined as the difference between the distance $d_{j'i}$ of the first neighbor not hydrogen bonded to $i$ (with label \textbf{5} in the figure), and the
distance $d_{j''i}$ of the last neighbor hydrogen bonded to $i$ (labeled \textbf{4}). As we will show, the $S$ state having high translational symmetry,
is characterized by large values of $\zeta$, with a clear separation between first and second shell (for example when the fifth molecule is in position \textbf{5} in Fig.~\ref{fig:1}A).
But these structures should not be confused with local crystalline structures, as they generally lack orientational order,
i.e., the neighbors in the second shell are not oriented according to the crystal directions (with well defined eclipsed and
staggered configurations), due to
their embedding in water's disordered network.
In the $\rho$ state
the second shell is collapsed, with a distribution of $\zeta$ values roughly centered around $\zeta=0$ (in Fig.~\ref{fig:1}A this state
is obtained for example when the fifth neighbor is in position \textbf{5'}) and comprising many configurations with
negative values of $\zeta$, resulting from the penetration of the first shell from an oxygen belonging to a distinct tetrahedra.

\subsection*{Two state model of water.\ }
At ordinary thermodynamic conditions, these two states are mixed and the
free energy ($G$) of the mixture takes the form of a regular solution, i.e., the simplest non-ideal model
of liquid mixture \cite{tanaka2000thermodynamic}. 
\begin{equation}\label{eqn:two_state}
G = G_\rho+s \Delta G+k_BT\left[s\log s+(1-s)\log (1-s)\right]+Js(1-s),
\end{equation}
where $s$ is the fraction of the $S$ state, $G_\alpha$ ($\alpha=\rho,S$) is the free energy of the pure component,
$\Delta G=G_S-G_\rho$, and $J$ is the coupling between the two states, i.e., the source of non-ideality.
Unlike ordinary regular solutions, the fraction $s$
is not fixed externally, but by the equilibrium of the conversion reaction between the two
states, $\rho\leftrightarrows S$, obtained by equating their chemical
potentials ($\mu_S=\mu_\rho$)
\begin{equation}\label{eqn:s_value}
 \Delta G+k_BT\log\left(\frac{s}{1-s}\right)+J(1-2s)=0. 
\end{equation}
If $J>0$ the mixing is endothermic, and the model displays a critical point at $k_BT_c=J/2$, below which the
system undergoes a liquid-liquid demixing transition.

\subsection*{Simulations.} To test our model we conducted molecular dynamics simulations of the TIP4P/2005 model of water, one of the best
models of the liquid state~\cite{vega_tip4p2005_criticalpoint}. Simulations were run
for many state points covering a large area of the liquid state, with temperatures ranging from $T=200$ K to $T=350$ K,
and pressures ranging from $P=-1$ kbar to $P=3$ kbar. To identify hydrogen bonds we adopted the definition found in Ref.~\cite{luzar_chandler}
and widely adopted in simulations of liquid water. Temperature is always expressed in K, pressure in bar, distance in
nm, density $\rho$ in g/cm$^3$, and isothermal compressibility $\kappa_T$ in bar$^{-1}$. For more details on the simulation methods refer to the Methods section.

\subsection*{Relevance of the two-state picture: A strong support from simulations. }

\begin{figure}[ht]
 \centering
 \includegraphics[width=8cm]{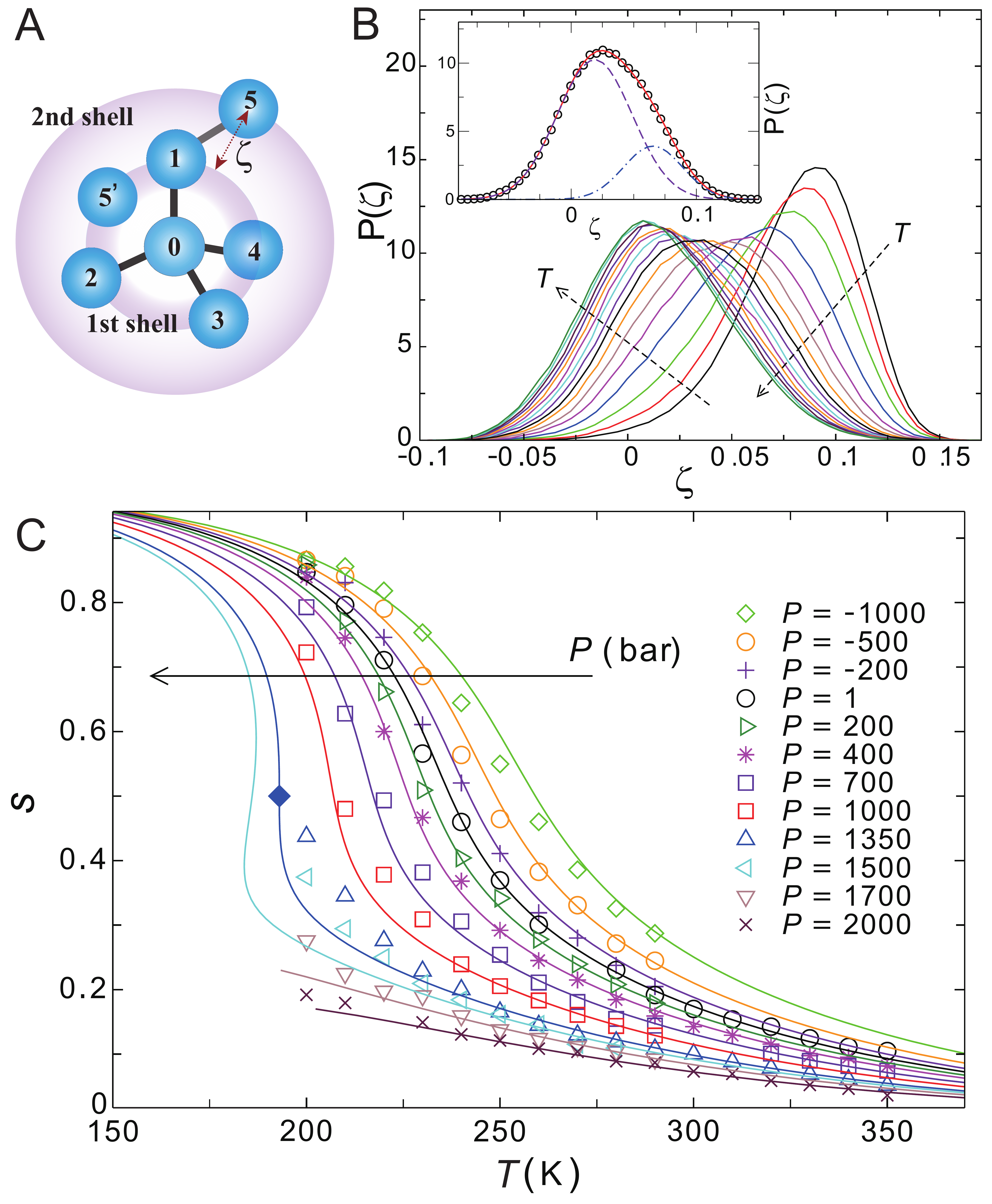}
 \caption{$T$, $P$-dependence of the structural order parameter. 
 (\textit{A}) Schematic representation of the local environment around a water molecule (labeled \textbf{0}), showing the
 position of the hydrogen-bonded molecules that form the first shell (with labels from \textbf{1} to \textbf{4}). Also
 depicted is the closest non-hydrogen-bonded molecule for two different states: the S state, when the fifth neighbor is in position
 \textbf{5}, so that there is a clear separation between first and second shell, with the order parameter $\zeta$ representing the
 distance between them; the $\rho$ state, with the fifth neighbor in position \textbf{5'} so that the second shell is collapsed onto the first one.
 (\textit{B}) Distribution function of the order parameter $\zeta$ at ambient pressure ($P=1$ bar) and at different
 temperatures. The temperatures range from $T=200$ K to $T=350$ K in steps of $10$ K (the arrows indicate increasing
 temperatures). The inset shows the distribution function at $T=280$ K obtained from the simulations (circle symbols),
 and the decomposition in two Gaussian populations: the $\rho$ population in dashed line, the $S$ population in dashed-dotted line,
 and the sum of both populations in the continuous (red) line.
 (\textit{c}) Values of the fraction of the $S$ state ($s$) as a function of temperature for all simulated pressures.
 The symbols are the values obtained by the decomposition of the order parameter distribution, $P(\zeta)$,
 at the corresponding state point. Continuous lines are fits according to the two-state model,
 with the following parameters defined in the text: $a_1=-2.90\cdot 10^{2}$, $a_2=-9.00\cdot 10^{1}$, $a_{11}=-6.32\cdot 10^{2}$,
 $a_{12}=-1.23\cdot 10^{2}$ and $a_{22}=-1.86\cdot 10^{1}$.
 The full diamond shows the location of the critical point, $T_c=193$ K, $P_c=1350$ bar~\cite{vega_tip4p2005_criticalpoint}.}
 \label{fig:1}
\end{figure}

Figure~\ref{fig:1}B shows the probability distribution of the order parameter $\zeta$ at ambient pressure ($P=1$ bar) and
for different temperatures, ranging from $T=200$ K to $T=350$ K. The change of the distribution shows
a remarkable non-monotonic change (of both height and width) with temperature. This behavior can be rationalized by
decomposing the distribution function in two gaussian populations, each varying monotonically with temperature (see Methods for the details). An example of this decomposition is shown
in the inset of Fig.~\ref{fig:1}B for the distribution function at $T=280$ K, which is close to the state point of density
maximum.  The $\rho$ population (dashed line in the inset of Fig.~\ref{fig:1}B) is centered in proximity of $\zeta=0$, meaning that there
is a large fraction of configurations in which the first shell (defined as the hydrogen bonded molecules to the central molecule)
is being penetrated by oxygen atoms belonging to different tetrahedral units. The $S$ population is instead characterized by high values
of $\zeta$, having a well formed second shell. The fitting procedure produces a reliable decomposition for state points having
$s<70\%$ ($s$ is the fraction of the $S$ state in the model of Eq.~(\ref{eqn:two_state})), where the fitting parameters are well behaved.
This covers the whole region of the phase diagram accessible to experiments. But for deeply supercooled states at low pressures (below
$T\approx 230$ K at ambient pressure), the fraction of the $\rho$ state becomes small, and the estimation of $s$ from unconstrained
fits becomes more difficult (see Methods). Nonetheless, as we will show later, the predictions of the model for the deeply supercooled states
are still in good agreement with simulations.

\subsection*{Dependence of the structural order parameter on temperature and pressure. }
From the decomposition we can extract the fraction of the $S$ state in the liquid, i.e., the parameter $s$ in the two-state model
of Eq.~(\ref{eqn:two_state}). The values of $s$ extracted from all simulated state points are shown in Fig.~\ref{fig:1}C.
For any pressure, the fraction of the $S$ state increases monotonically by lowering the temperature, and the increase is
steeper at lower pressures. At low temperatures, we can see a big jump in the values of $s$ between $P=1000$ bar and
$P=1350$ bar. This signals the close presence of a liquid-liquid critical point. In fact, previous studies of TIP4P/2005 water (and with the same system size)
have determined its location at $T_c=193$ K and $P_c=1350$ bar~\cite{vega_tip4p2005_criticalpoint}, 
even if the exact location of the critical point is still ambiguous \cite{patey2013}. 
Next we fit the two-state model of Eq.~(\ref{eqn:s_value}) to our simulation results (see Methods for the details of the fitting). 
The results of the fit are shown as continuous lines in Fig.~\ref{fig:1}C, demonstrating that the two-state model provides a
very good representation of the results obtained from simulations (represented by symbols in the same figure). Note that the agreement holds
to very high temperatures, and for all pressures, suggesting a strong microscopic basis for the relevance of our order parameter. 
This is in stark contrast to previous attempts to obtain a two-state description of water
from microscopic information~\cite{appignanesi,poole_mixture,wikfeldt_bimodal}, where the agreement
was restricted to the deeply supercooled region. 
We note that most of previous models have unphysical saturations of the value of $s$ at high temperature around 0.5 or even higher, 
unlike our model where $s \ll 1$ at high temperatures (see also Ref. \cite{tanaka_review} for a review). 
It is also worth mentioning that the only region of significant discrepancy is limited around the critical point, which is due not only to the higher uncertainty
in accessing $s$ around the critical region, but possibly also to critical fluctuations which are not incorporated in the
present mean-field two-state model (but which is possible with crossover theory~\cite{holten_anisimov}).

\subsection*{Two-state description of water anomalies.}

\begin{figure}[ht]
 \centering
 \includegraphics[width=8cm]{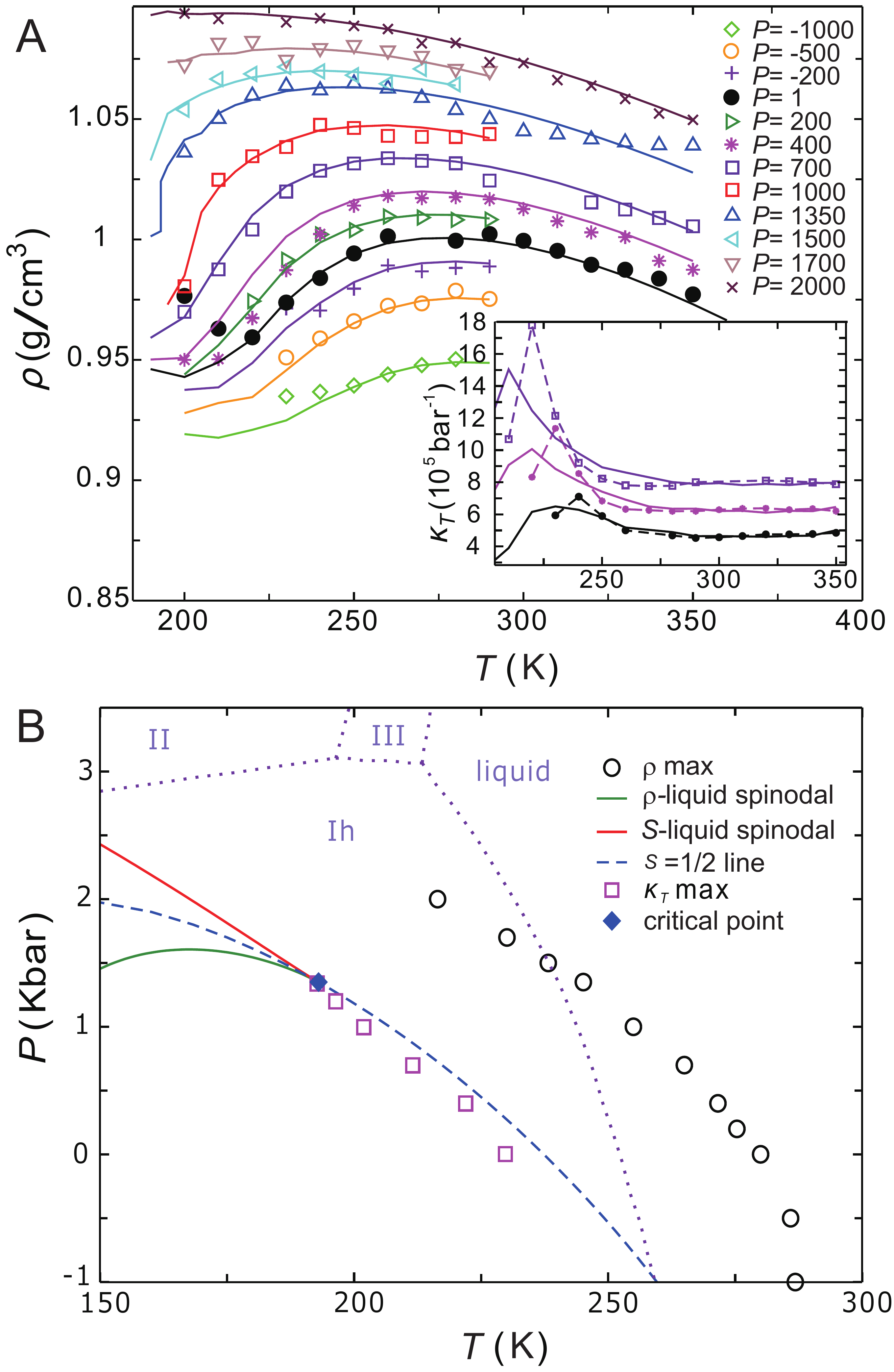}
 \caption{Thermodynamic and phase behavior of water. 
(\textit{A}) Temperature dependence of density for several pressures. Continuous lines are simulation results, while symbols are obtained from the two-state model.
 The inset shows the compressibility for $P=1,400,700$ bar: dashed lines are simulation results, while symbols are two-state model predictions. The curves in the inset are traslated
 on the $y$-axis to improve readability.
(\textit{B}) Phase diagram of the two-state model. The Widom line (dashed line) is by definition the line where $s=1/2$. The compressibility
maximum (squares) from simulations lies close to the Widom line, eventually converging at the critical point (full diamond). Below
the critical point the two components undergo macroscopic phase separation. The spinodals of the transition are denoted with full
lines: green for the $\rho$ liquid, and red for the $S$ liquid.
Circles show the locus of density maximum, which continuously
increases as water is stretched, in agreement with experiments at extreme negative pressures~\cite{caupin}.
Dotted lines show the location of the coexistence lines between the liquid and different forms of ice taken from Ref.~\cite{vega_tip4p2005_phasediagram}
 }
 \label{fig:2}
\end{figure}

We will now check whether the two-state picture extracted so far can account for water's anomalies. From the free energy of Eq.~(\ref{eqn:two_state}) it is possible
to derive the anomalous contribution to density, which takes the form $\rho=N/V$ with $V=V_\rho+s \Delta V$,
where $V_\rho$ is the volume of $N$ molecules of the pure $\rho$-state and $\Delta V$ is the volume difference between $S$-state and $\rho$-state. 
On the $T$-$P$ dependence of $\Delta V$ and $V_\rho$, see Methods. 
Figure~\ref{fig:2}A shows density isobars measured in simulations (continuous lines) which are compared to
the results from the two-state model (symbols). We can see that the two-state model correctly represents the intensity and location of the density anomaly
for all studied pressures (with the anomaly disappearing at high pressures). The agreement is more remarkable in the relevant region of the anomalies,
while it is approximate at very low temperatures (for which we know the estimation of $s$ being subject to higher uncertainty). For example the model
predicts a density minimum at around $220$ K at ambient pressure (full circles in Fig.~\ref{fig:2}A), which is found instead at $200$ K in simulations.
The inset shows the isothermal compressibility $\kappa_T$ for pressures $P=1,400,700$ bar. As in the main figure, continuous lines are direct simulation
results, showing the rapid increase at low temperatures, while symbols are predictions from the model (see Methods).
Isothermal compressibility anomalies are much harder to describe accurately, both because compressibility is a second derivative of the model's free energy, $G$  
(see Methods), 
thus suffering bigger uncertainty, and also because its anomaly is located at very low temperatures, where it is more difficult to get
reliable estimation of the fraction $s$. Nonetheless, as shown in the inset of Fig.~\ref{fig:2}A, the model predicts within $10$ K the location
of the compressibility maximum, and also its increased intensity at higher pressures (eventually diverging at the critical point). 

In the Appendix we show that, with minor 
approximations, it is possible to obtain the fraction of the $S$ state not only from the distribution
of $\zeta$ but also directly from the experimentally measurable O-O radial distribution function. 
This opens up a possibility to estimate the structural order parameter, i.e., the degree of translational order in the second shell, from scattering experiments.

\subsection*{Phase behavior of water.}
We summarize the phase behavior of the two-state model in the phase diagram of Fig.~\ref{fig:2}B. The Widom line of
the model, where $s=1/2$, lies close to the compressibility maximum line obtained from simulations (open squares), with the two lines converging
at the critical point. 
Here we note that the location of the Widom line is determined by the condition $\Delta G(T,P)=0$, 
i.e., the two-state feature without cooperativity (see Eq. (\ref{eqn:s_value})). This indicates that the isothermal compressibility anomalies 
in this temperature range are not due to critical phenomena associated with LLCP, but due to the sigmoidal change in $s$ 
characteristic of the two-state model (Schottky-type anomaly). 
Also shown in the phase diagram are the spinodals (or, stability limits) of the $S$ liquid and the $\rho$ liquid,
which in the liquid-liquid critical point scenario are usually called the low density liquid (\emph{ldl}) and the high density liquid (\emph{hdl}) respectively.

\subsection*{Microscopic structural features of locally favored structures.}
\begin{figure*}[ht]
 \centering
 \includegraphics[width=15cm]{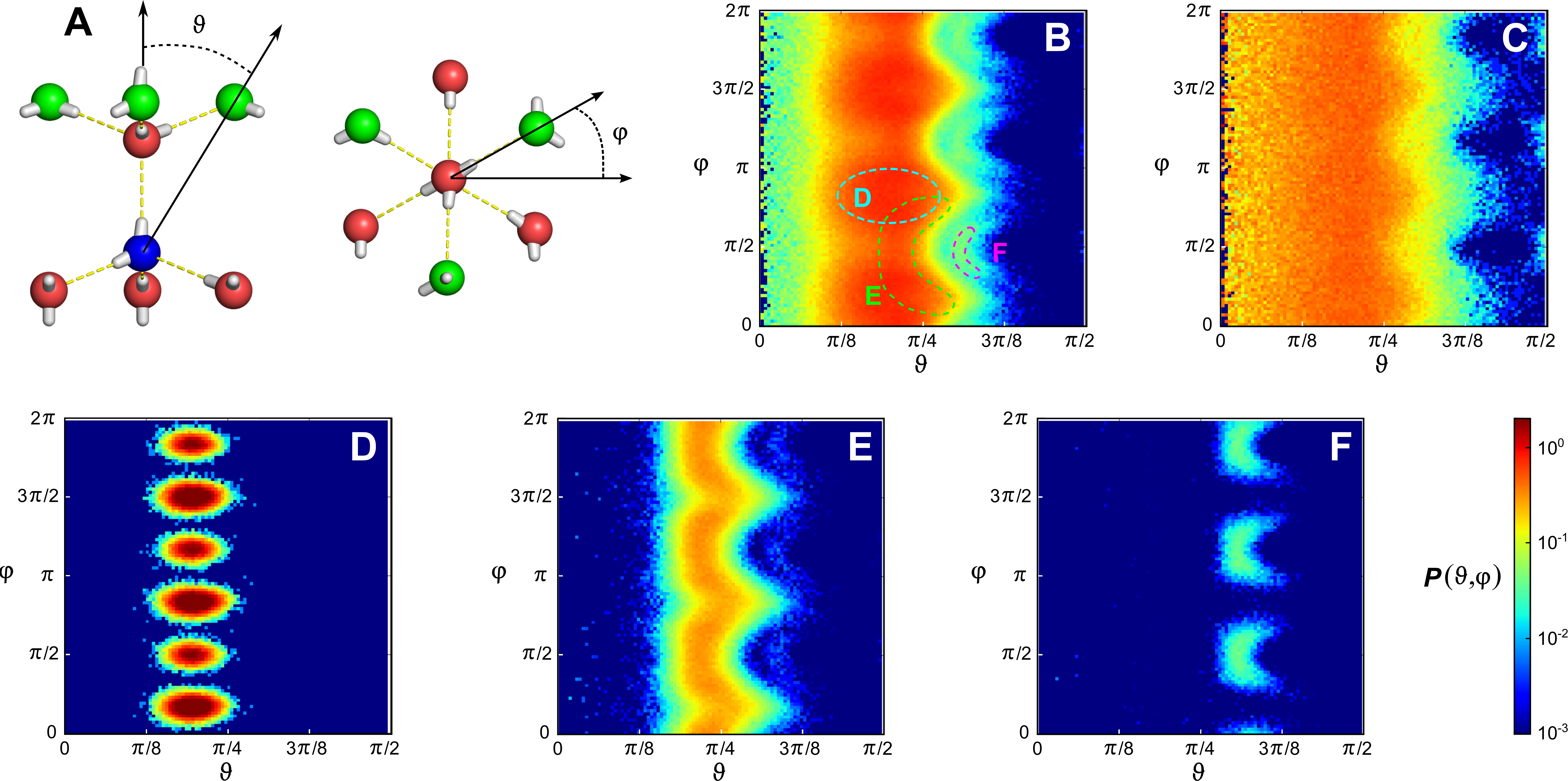}
 \caption{Structural characterization of the $S$ state. 
(\textit{A}) Polar maps of the relative position of second nearest neighbors (green oxygens) of a given molecule (blue oxygen), with
 its first nearest neighbors (red oxygens) oriented so to minimize the mean square distance from a regular tetrahedron (see text for details).
(\textit{B}) Liquid state at $T=200K$ and $P=1$ bar. Only particles which are not identified as crystalline, and with $\zeta\in[0.075,0.1125]$
 are plotted. The dashed labeled lines are guides to the structures into which the $S$ state can be decomposed, and which are plotted in the corresponding panels.
(\textit{C}) Same as in panel \emph{b}, but for the state point $T=280$ K and $P=1$ bar.
(\textit{D}) Hexagonal Ice ($I_h$) at $T=200$ K and $P=1$ bar.
(\textit{F}) Liquid state at $T=200$ K and $P=1$ bar, where only particles belonging to 4-membered rings are plotted.
 The color bar represents the probability density for panels (B), (C), (E), and (F), which are all computed from the same set of configurations.
 }
 \label{fig:S_structure}
\end{figure*}

We now investigate the microscopic features of locally favored states. We have shown that the
$S$ state can be identified with configurations having a high degree of translational order, meaning that
second nearest neighbors are at approximately the same distance from the central oxygen atom. We now discuss the
orientational order of second nearest neighbors. Crystalline configurations are
characterized by full orientational order, with second nearest neighbors occupying the characteristic eclipsed
and staggered orientations present in the stable ice $I_c$ and $I_h$ polymorphs (see Fig.~\ref{fig:S_structure}A). To investigate the structure
of the $S$ state, we consider only oxygens atoms for which our order parameter $\zeta$ is in the range $\zeta\in[0.075,0.1125]$,
where the $S$ population peaks (see Fig.~\ref{fig:1}). We also exclude from the analysis particles which
are identified as belonging to small crystals that spontaneously form and dissolve in a supercooled melt. To identify
crystalline particles we use standard order parameters based on Steinhardt rotational invariants~\cite{steinhardt,reinhardt2012local}.
For each of the oxygens atoms previously defined, we determine the
optimal rotation that minimizes the following root mean square deviation $\sqrt{\sum_{i=0}^4(\mathbf{r}_i-\mathbf{v}_i)^2}$,
where $\mathbf{r}_i$ ($i=1..4$) are the unitary vectors joining the central oxygen atom ($\mathbf{r}_0$) with its four nearest neighbors,
and $\mathbf{v}_i$ ($i=1 \cdots 4$) are the directions of a reference tetrahedron, given by
$\mathbf v_0=(0,0,0)$, $\mathbf v_1=(\sqrt{2/3},-\sqrt{2}/3,-1/3)$, $\mathbf v_2=(-\sqrt{2/3},-\sqrt{2}/3,-1/3)$,
$\mathbf v_3=(0,2\sqrt{2}/3,-1/3)$ and $\mathbf v_4=(0,0,1)$. The same rotation is applied to second nearest-neighbors,
defined as all oxygen atoms whose distance from first nearest neighbors is within $1.2$ times the average oxygen-oxygen distance.
We then compute the probability distribution for the position of second nearest neighbors in spherical coordinates, according to the usual
transformations: $r=\sqrt{x^2+y^2+z^2}$, $\theta=\cos^{-1}(z/r)$ and $\phi=\tan^{-1}(y/x)$, where $z$ is the axis connecting the central
oxygen atom with the closest vertex $\mathbf{v}_i$ of the regular tetrahedron.
A schematic representation of the coordinate system is shown in Fig.~\ref{fig:S_structure}A.
The probability to find a crystalline particle with hexagonal planes pointing in the $(\theta+d\theta,\phi+d\phi)$ direction is then given by
$P(\theta,\phi)\sin\theta\,d\theta\,d\phi$.

Typical results for equilibrium configurations are shown in Fig.~\ref{fig:S_structure}. To aid the understanding of these plots, we first report in Fig.~\ref{fig:S_structure}D the results
for the hexagonal ice crystal. The six well defined peaks correspond to the possible positions of second nearest neighbors
in the crystal, showing full orientational order: larger peaks correspond to staggered configurations, smaller peaks to eclipsed configurations.
In Fig.~\ref{fig:S_structure}B we plot the results for the $S$ state of liquid water at $T=200$ K and $P=1$ bar.
We first notice that the full orientational order found in the crystal is lost in the $S$ state. Nevertheless, one can still
identify structural patterns that characterize the $S$ state, and that are marked by the dashed lines in Fig.~\ref{fig:S_structure}B.
The first pattern (\textbf{D}) corresponds to staggered arrangements of molecules, providing strong evidence that the
$S$ state is a precursor of crystallization, i.e. that it is
along the microscopic pathway that water undergoes when transforming from liquid to solid.
The second prominent structural pattern found in the $S$ state is denoted by \textbf{E} in Fig.~\ref{fig:S_structure}B
and it is not found in stable ice crystals. This pattern is centered around eclipsed configurations, but it is characterized by
fluctuations which are distinct from the one found in the hexagonal crystal. While second nearest neighbors in crystalline configurations
are involved in loops of six hydrogen-bonded oxygen atoms, the pattern in \textbf{E} is instead due to five-membered rings. This is shown in Fig.~\ref{fig:S_structure}E,
where only oxygen atoms belonging to
five-membered rings are plotted, displaying the same pattern found in the $S$ state. 

\subsection*{Roles of locally favored structures in ice crystallization and its avoidance.}
Pentagonal rings, loops of five water molecules bonded to each other
through hydrogen bonding, thus act as the source of frustration against crystallization \cite{TanakaGJPCM,ShintaniNP}. In order to crystallize, the $S$ structure
needs first to break an hydrogen bond and then orient its neighbors along the crystal's directions. The $S$ state is energetically
stable with respect to the disordered $\rho$ state (since each pentagon ring adds one hydrogen bond
to the structure), but pays a high entropic cost, due to the missing degrees of freedom when closing a ring. Pentagon rings could then be responsible for the
high degree of supercooling reachable with water, stabilizing the $S$ state and frustrating the crystallization transition.
Finally, the structures denoted by \textbf{F}
represent four-member rings, which are plotted in Fig.~\ref{fig:S_structure}F, and are present in far less extent than five-member rings. 
Thus, the $S$-state is characterized by mixed structural signatures, one of which is consistent with the crystal structure and the other 
is not.

\begin{figure}[ht]
 \centering
 \includegraphics[width=6cm]{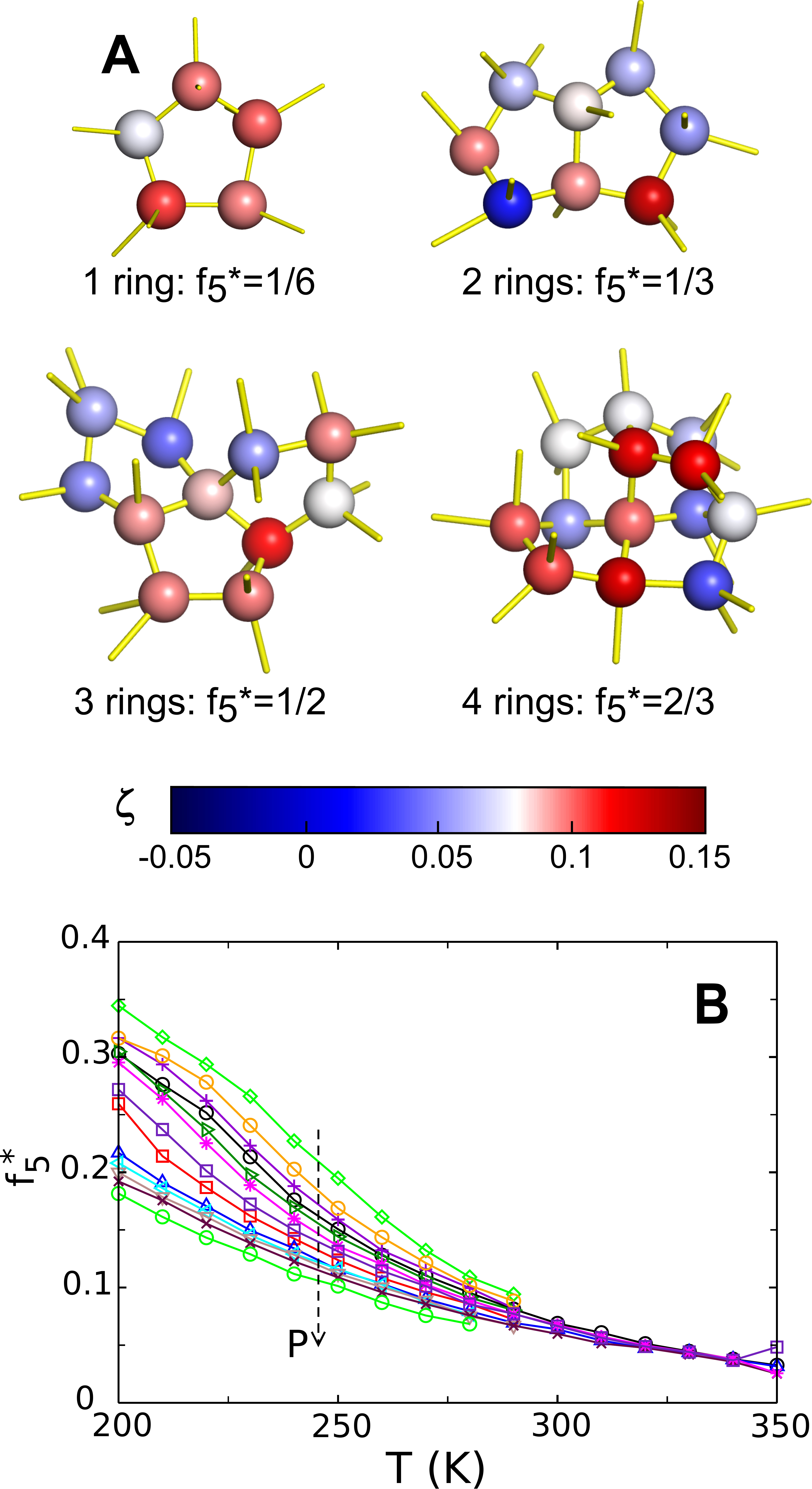}
 \caption{Pentagonal rings. (\textit{A}) Snapshots of water configurations with respectively
 one, two, three and four five-membered rings. Oxygen atoms are colored according to the value
 of the translational order parameter $\zeta$. (\textit{B}) Fraction of pentagonal rings in the $S$ structures, as a function of $T$ for all pressures
 considered in the work. The order of increasing pressure is indicated by the dashed arrow.}
 \label{fig:rings}
\end{figure}

We investigate the statistics of five-membered rings in Fig.~\ref{fig:rings}.
The first panel shows some snapshots of pentagonal rings for configurations having respectively one, two, three and four five-membered rings.
For each $S$ structure the maximum number of pentagonal rings is six. In Fig.~\ref{fig:rings}B we show how the fraction of pentagonal
rings ($f_5^*$) in the $S$ structures changes with temperature, for different values of the pressure. The number of five-membered rings increases with decreasing
temperature and pressure. A decrease in temperature increases the population of the $S$ state, so favoring crystallization, but this
is partly counterbalanced by an increase five-membered rings.

\section*{Conclusion}
To conclude, we have provided a microscopic description of water's anomalies
based on locally favored states defined as structures with local translational order.
Microscopically, these states reflect the underlying crystallization behavior
of water. Due to its strong directional bonding,
the crystallization pathway of water can be approximated in two steps: in the first step, water develops translational order, which is reflected
in the population of $S$ states; in the second step, it develops orientational order where the hydrogen bonds in the shell of second
nearest neighbor acquire the staggered and eclipsed configurations which characterize the hexagonal and cubic forms of ice.
The $S$-state is a locally favored state (energetically stabilized) with a high abundance of pentagonal rings, loops of five particles bonded to each other
through hydrogen bonding. Five-membered rings increase with lowering the temperature and thus act as a source of frustration against crystallization \cite{TanakaGJPCM,ShintaniNP}.
The model which results from this framework is compatible with
water's anomalies over a wide range of temperatures and pressures. 

Moreover it can be fully consistent with the
liquid-liquid critical point scenario, where the $ldl$ and $hdl$ states are respectively the $S$-dominant and $\rho$-dominant states below
the critical point. In fact, the order parameter $\zeta$ includes the local configurations responsible for
water's two-lengthscales average interaction~\cite{mishima_stanley}. 
We note that, within the two-state model, 
the presence of a critical point associated with demixing is not a necessary
condition for the existence of thermodynamic anomalies, and the anomalies survive even if the mixture lacks a critical point
($J=0$ in Eq.~(\ref{eqn:two_state})) \cite{tanaka2000thermodynamic}. Such a possibility has recently been suggested for TIP4P/2005 water \cite{patey2013}. 
We stress that the parameters of the model were obtained only from microscopic information, and then its predictions were compared to
the anomalies of water. It would be possible also to use the model in a phenomenological way, by fitting the anomalies to improve the model parameters,
for example improving the estimates in the deeply supercooled region (obtaining better estimates for the density minimum and the isothermal compressibility
maximum). In this work we have avoided such an approach to show that a two-state description of the phase behavior of
water is possible from microscopic information. Once this is confirmed, precision fitting of the anomalies can provide additional insights~\cite{tanaka2000simple,holten_anisimov}.

For this reason
we have also provided an approximate scheme to extract the model's parameters from experimentally accessible measurements (see Appendix). Knowledge
of the oxygen-oxygen radial distribution function is in fact accessible with both neutron and X-ray scattering methods 
(see, e.g., Ref. \cite{Nilsson2011}). Performing these
measurements over a wide range of temperature and pressures could provide important information, such as the nature of the non-idealities of the mixture
(for example whether they have an energetic or entropic origin~\cite{holten_anisimov}). 

Here it may be worth noting the roles of locally favored structures in crystallization. 
We show that locally favored structures in water not only have translational order in the second shell, but also contain five-membered rings of hydrogen-bonded molecules, 
indicating a mixed character of their roles in crystallization: the former helps crystallization, whereas the latter causes frustration against crystallization. 
This frustration effect may be related with the rather large degree of supercooling of water before homogeneous crystal nucleation of hexagonal or cubic ices takes place.

Finally, the validity of two-state models can be assessed
not only in experimental studies of pure water, but also in water mixtures, where a liquid-liquid transition of the water component has been observed~\cite{murata},
or in water-salt solutions, where the effect of salt on the structure of water is similar to the effect of pressure~\cite{Leberman1995,kobayashi2011possible}. 
We can further speculate that the two-state model based on the translational order of the second shell may be relevant also to other tetrahedral liquids 
(Si, Ge, silica, germania) \cite{TanakaWPRB,tanaka_review}, which play a crucial role in materials science.  




\section*{Methods}

\noindent
{\bf Molecular dynamics simulations.}
Molecular dynamics simulations were run using the Gromacs (v.4.5) molecular dynamics simulation package.
The isothermal-isobaric $NPT$ ensemble was sampled through a Nos\'e-Hoover thermostat and an isotropic Parrinello-Rahman barostat.
Lennard-Jones interactions have a cutoff at $0.95$ nm, and cutoff corrections are applied to both energy and pressure.
Electrostatic interactions are calculated through Ewald summations, with the real part being truncated at $0.95$ nm, and the
reciprocal part evaluated using the particle mesh method. The systems consist of $512$ molecules of water, and the total
simulation time varied from $200$ ns for the high temperature simulations, up to $1$ $\mu$s for the lowest temperatures.
The water force field is TIP4P/2005~\cite{vega_tip4p2005_criticalpoint}. Hydrogen bonds are located through geometric constraints on the relative positions
of the donor (D), acceptor (A) and hydrogen (H) atom. Two oxygen atoms are considered hydrogen bonded if their distance is within $0.35$ nm,
and the angle HDA is less than $30^\circ$~\cite{luzar_chandler}.

\noindent
{\bf Analysis of the distribution function $P(\zeta)$.}
The distribution function $P(\zeta)$ is decomposed in two gaussian populations. 
In order to distinguish unambiguously two gaussian populations with large overlap, we take into account the fact
that one of the two populations (corresponding to the $S$ state) should be characterized by fully formed translational
order up to second shell, so its distribution should be vanishingly small for $\zeta\rightarrow 0^+$. 
This means that the $S$ state is characterized by good translational order on both first and second shells.
Imposing this constraint on the fitting we are able to decompose the distribution in two populations for a large region of the phase
diagram. The fitting function satisfying the above constraint is given by 
\begin{eqnarray}
 P(\zeta)= & \frac{P(0)}{\exp({-\frac{m_\rho^2}{2\sigma_\rho^2}})}\exp\left({-\frac{(\zeta-m_\rho)^2}{2\sigma_\rho^2}}\right) + 
\nonumber \\
   & +\left(1-\frac{\sigma_\rho\sqrt{2\pi}P(0)}{\exp({-\frac{m_\rho^2}{2\sigma_\rho^2}})}\right)\frac{\exp\left({-\frac{(\zeta-m_S)^2}{2\sigma_S^2}}\right)}{\sigma_S\sqrt{2\pi}}, \nonumber
\end{eqnarray}
where $m_\rho,\sigma_\rho,m_S,\sigma_S$ are the fitting parameters. Both $m_\rho$ and $m_S$ are monotonically decreasing with temperature,
showing that states become more and more structured at low temperatures, while $\sigma_\rho$ and $\sigma_S$ are increasing with temperature,
as expected by the increase of thermal fluctuations.
The fitting performs well for state points characterized by $s\lesssim 0.7$ (so throughout the experimentally accessible region), but
at very low temperatures and pressures the fraction of the $\rho$ state becomes small, increasing the uncertainty of the fit. One possible
way to overcome these difficulties is by noting that the $\rho$ fluid should behave like a fluid without anomalies, and thus its parameters
($m_\rho$ and $\sigma_\rho$) should be well behaved functions of $P$ and $T$. We then choose to estimate the value of $m_\rho$ with a quadratic
extrapolation from the values obtained at state points with $s\lesssim 0.7$. This procedure degrades the accuracy of the model only at the lowest
temperatures and for small pressures (as can be seen in the estimates for the isothermal compressibility maximum and the minimum in density in Fig.~\ref{fig:2}A),
but the agreement with simulations is still satisfactory.

\noindent
{\bf Fittings of the $T$-$P$ dependence of the order parameter, density, and isothermal compressibility.}
The two-state model predictions are based on the free energy expression given by Eq.~(\ref{eqn:two_state}).
To fully specify the model, an expression for the free energy difference between the two bulk states, $\Delta G$,
has to be provided. In our model this difference is expressed as a second order expansion around the known location
of the liquid-liquid critical point:
\begin{equation}
-\Delta G=a_1\hat{T}+a_2\hat{P}+a_{11}\hat{T}^2+a_{12}\hat{T}\hat{P}+a_{22}\hat{P}^2, \nonumber
\end{equation}
where $\hat{T}=(T-T_c)/T_c$ and $\hat{P}=(P-P_c)/P_c$. 
We employ $T_c=193$ K and $P_c=1350$ bar, which are reported for the same system~\cite{vega_tip4p2005_criticalpoint}. 
This also fixes the value of $J=2k_{\rm B}T_c$. The coefficients of the expansion $a_{\alpha\beta}$ are given in the caption of Fig.~\ref{fig:1}.
The resulting free energy allows the calculation of the anomalous term of any thermodynamic property $A$, i.e. the
difference between the value of $A$ in the mixture and its value in a pure state (the background), $A_\rho$ or $A_S$.
We define the anomalous term, $\Delta A$, with the following expression: $A=A_\rho+s\Delta A$. For example,
the expressions for the density and compressibility anomalies are found respectively with first and second order derivatives
of the free energy with respect to pressure
\begin{eqnarray}
&\Delta V = \frac{\partial\Delta G}{\partial P} = \frac{-a_2+a_{12}\hat{T}-2a_{22}\hat{P}}{P_c}, \nonumber \\
&\Delta\kappa_T = -\frac{1}{V}\frac{\partial (V-V_\rho)}{\partial P} = \frac{\Delta V^2}{Vk_BT/(s(1-s))-2JV}+\frac{2a_{22}}{VP_c^2}. \nonumber
\end{eqnarray}

The calculation of the absolute value of the thermodynamic properties requires a reasonable assumption on the
 $T$-$P$ dependence of the background part, which is usually written as a low-order polynomial. For example
 $V_\rho=b_0+b_1\,\hat{T}+b_2\,\hat{T}^2$. We stress that the success of a two-state model depends on the ability to
 determine the location and intensity of the anomalies, which do not depend on the knowledge of the background terms.

 \section{Appendix: radial distribution functions}

The microscopic approach described in the main text requires the knowledge of the instantaneous positions of both oxygen and hydrogen
atoms in the system, thus being suitable only for computational studies. Here we show that, with minor
approximations, it is possible to obtain the fraction of the $S$ state not only from the distribution
of $\zeta$ but also from experimentally measurable quantities, which can probe the degree of translational order. The definition of $\zeta$ involves the difference
between the distance of the first non-hydrogen bonded oxygen and the last hydrogen bonded oxygen. The hydrogen
bond interaction is a limited range interaction, and breaks when the oxygens involved in the bond are too
far apart. We call this cutoff distance $r_{\rm H-bond}$. In our simulation study we have set this distance at
$r_{\rm H-bond}=0.35$ nm~\cite{luzar_chandler}.  This means that if
we look at two oxygen atoms whose distance is about $r_{\rm H-bond}$, these atoms (with high probability) do not share
an hydrogen bond. At the same time we know that in the $S$ state the first non-hydrogen bonded atom has a vanishingly
small probability to be at distances close to the first neighbors' shell ($\zeta=0$). It follows that the contribution
to the radial distribution function at distance $r_{\rm H-bond}$ comes predominantly from the $\rho$ state. To clarify
this point, we consider the radial distribution functions of oxygen atoms, $g(r)$, plotted in Fig. S1a. Continuous lines show
the radial distribution function for different state points along the Widom line (where by definition $s=1/2$):
the distributions are all very close to each other, with an isosbestic point approximately at $r_{\rm H-bond}=0.35$ nm. 
The reason for the isosbestic point is that, as noted above, the value of the radial distribution function at
$r_{\rm H-bond}$ is proportional to the fraction of the $\rho$ state, and this is constant at the Widom line.
Coincidentally, the value of the radial distribution function at $r_{\rm H-bond}$
is approximately $g(r_{\rm H-bond})=0.5$, suggesting a direct proportionality between the fraction of the $\rho$ state and $g(r_{\rm H-bond})$.
This is confirmed by plotting the radial distribution functions of state points with a low fraction of the $\rho$ state
($P=-1000$ bar, $T=200$ K, dotted-dashed line) and a high fraction of $\rho$ state ($P=2000$ bar, $T=350$ K, dashed line),
where the value of $g(r_{\rm H-bond})$ is found to be close to $0$ and $1$ respectively. We thus write the following relation between
the height of $g(r)$ and $s$ (fraction of the $S$ state): $s\cong 1-g(r_{\rm H-bond})$.
Figure S1b compares the results of this relation with the results obtained by fitting the distribution
function of the order parameter $\zeta$. In the inset these two values are plotted for all state points considered in
this work, showing that indeed the relation $s=1-g(r_{\rm H-bond})$ holds to a good approximation. The main panel
in Fig. S1b shows the temperature dependence of these two values, showing again the good agreement between
the two calculation methods. Note that the relation $s=1-g(r_{\rm H-bond})$ seems to hold better for $s<0.5$, which is the same
range that the experiments are able to access.

\begin{figure}[h!]
 \centering
 \includegraphics[width=7cm]{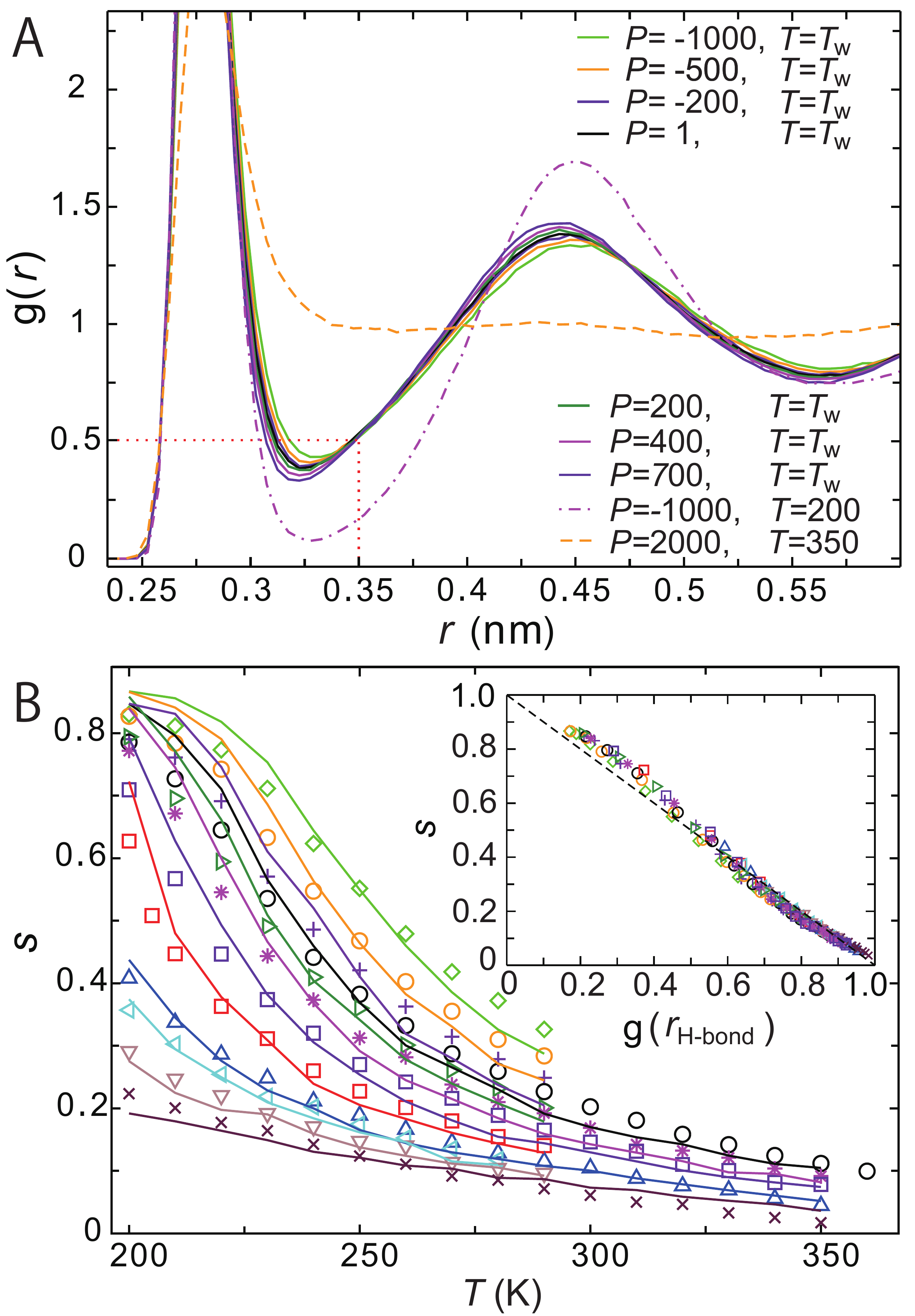}
 \caption{Estimation of $s$ from the radial distribution function of oxygen atoms, $g(r)$. 
(\textit{A}) Radial distribution function $g(r)$ for state points along the Widom line (continuous lines). The functions
were obtained by linear interpolation of radial distribution functions of the closest simulated state points.
Also plotted are the radial distribution functions of two state points with respectively a high fraction of the $S$ state
($P=-1000$ bar, $T=200$ K, dotted-dashed line) and a high fraction of $\rho$ state ($P=2000$ bar, $T=350$ K, dashed line)
The dotted lines denote the isosbestic point close to $r_{\rm H-bond}=0.35$.
(\textit{B}) Comparison between the values of $s$: symbols represent values obtained from
the relation $s\cong 1-g(r_{\rm H-bond})$, while lines are the values obtained from the distribution function of the order
parameter $\zeta$ (the lines in this figure represent the same data as the symbols in Fig. 1c).
The inset shows the comparison between the value of $s$ obtained from the distribution function of the order
parameter $\zeta$ and the value of $g(r_{\rm H-bond})$ at the same state points. 
}
 \label{fig:3}
\end{figure}

\section*{Acknowledgements}
This work was partially supported by a Grant-in-Aid for Scientific Research (S) from JSPS,  
Aihara Project, the FIRST program from JSPS, initiated by CSTP, and a JSPS Postdoctoral Fellowship.


\end{document}